\documentclass{aa}
\usepackage{graphicx,amssymb,natbib}
             
\newcommand{\Teff}{$T_{eff}$}
\newcommand{\lgg}{log~$g$}
\newcommand{\Vsini}{$V\sin i$}
\newcommand{\kms}{km~s$^{-1}$}
\newcommand{\header}[1]{\multicolumn{1}{c}{\textrm{#1}}}

\begin{document}
\title{Search for broad absorption lines in spectra of stars in the field of 
supernova remnant RX J0852.0-4622 (Vela Jr.)}
\author{
A. F. Iyudin\inst{1,2}
\and
Yu. V. Pakhomov\inst{3}
\and
N. N. Chugai\inst{3}
\and
J. Greiner\inst{2}
\and
M. Axelsson\inst{4}
\and
S. Larsson\inst{4}
\and
T. A. Ryabchikova\inst{3}
}

\institute{Max-Planck-Institut f\"ur extraterrestrische Physik, Postfach 1312, D-85741 
Garching, Germany
\and
Skobeltsyn Institute of Nuclear Physics, Moscow State University, 
Vorob'evy Gory, 119992 Moscow, Russian Federation
\and
Institute of Astronomy, RAS, Pyatnitskaya 48, 119017, Moscow, Russian Federation
\and 
Stockholm University, AlbaNova University Center, Department of Astronomy, 
SE-106 91 Stockholm, Sweden
}

\authorrunning{A.F.Iyudin et al.}
\titlerunning{Broad absorption lines in the field
of SNR Vela Jr.}

\date{Received ...;   Accepted ... }

\abstract {}
{Supernova remnant (SNR) RX J0852.0-4622 is one of the youngest and is most likely the closest among known galactic 
supernova remnants (SNRs). It was detected in X-rays, the $^{44}$Ti $\gamma$-line,
and radio.
We obtain and analyze medium-resolution spectra of 14 stars in the direction 
towards the SNR RX J0852.0-4622 in an attempt to detect broad absorption lines of 
unshocked ejecta against background stars.}
{Spectral synthesis is performed for all the stars in the 
wavelength range of 3740-4020\AA\ to extract the broad 
absorption lines of Ca~II related to the SNR RX J0852.0-4622.}
{We do not detect any broad absorption line and place a 3$\sigma$ upper limit on 
the relative depths of $<$0.04 for the broad Ca~II
absorption produced 
by the SNR. We detect narrow low and high velocity absorption components 
of Ca~II. High velocity $|V_{LSR}|\sim100-140$\kms\ components
 are attributed to radiative shocks in clouds engulfed by the old Vela SNR.
The upper limit to the absorption line strength combined with the width and 
flux of the $^{44}$Ti $\gamma$-ray line 1.16~MeV lead us to conclude that 
SNR RX J0852.0-4622 was probably produced by an energetic SN~Ic explosion.}
{}


\maketitle

%
%__________________________________________________

\section{Introduction}

Young supernova remnants (SNRs) at the deceleration stage generally 
provide us with an 
opportunity to probe the supernova (SN) ejecta not yet polluted by the
circumstellar matter (CSM). The ejecta composition and structure are  usually
studied by analyzing the X-ray spectra emanating from the reverse shock. In some 
young SNRs related to core-collapse SNe, e.g., Cas~A, the optical emission
of undecelerated ejecta clumps are observed when they penetrate 
the post-shock layer, in which case they are powered by slow radiative shocks.
Unshocked ejecta material is cold and does not radiate, thus remaining invisible
in emission. 

However, in rare cases the unshocked ejecta of SNR can be observed in resonance
absorption lines  against a background light source seen through the SNR shell. 
Among Galactic SNR, this method has been applied successfully only to 
SN~1006, where the ejecta was observed in the ultraviolet absorption lines
against the spectra of hot stars and QSOs \citep{1997ApJ...481..838H,
2005ApJ...624..189W, 2007MNRAS.381..771H}. There is also one extragalactic 
SNR detected in absorption lines: the remnant of SN~1885 in M~31. It was 
first detected by ground-based imaging in Fe~I 3860\AA\ band against 
the M~31 bulge \citep{1989ApJ...341L..55F} and afterwards by {\em HST}
imaging and spectroscopy \citep{1999ApJ...514..195F}. The optical spectrum of
this SNR contains strong Fe~I,  Ca~II, and Ca~I broad resonance absorption lines
produced by the unshocked ejecta.

\begin{table*}[!t]
\begin{minipage}{180mm}
\label{tab:list}
\centering
\caption{List of the observed stars} 
\begin{tabular}{llccclrccc}
\hline
\hline
\header{N}&\header{Star}&    RA     &     Decl &\header{$V$}&
\header{Sp}&\header{$B-V)$}& $p$ & \header{S/N} & \header{$\sigma$}\\
&          &\multicolumn{2}{c}{(2000.0)}&\header{(mag)}&     &\header{(mag)}
&& &\\
\hline
1 & HD75309   & 08 47 28.0& -46 27 04&  7.84& B2Ib/II  & 0.01& 0.79& 270&0.008\\
2 & HD75820   & 08 50 26.0& -46 14 53&  8.64& B9V      &-0.02& 0.27& 260&0.007\\
3 & HD75873   & 08 50 48.8& -46 18 36&  8.10& A3II/III & 0.38& 0.20& 160&0.010\\
4 & HD75955   & 08 51 26.0& -45 37 23&  7.73& B9V      &-0.01& 0.67& 240&0.005\\
5 & HD75968   & 08 51 32.8& -46 36 36&  8.14& B9III/IV &-0.12& 0.33& 200&0.006\\
6 & HD76060   & 08 52 02.4& -46 17 20&  7.88& B8IV/V   &-0.09& 0.02& 350&0.005\\
7 & HD76589   & 08 55 23.0& -46 53 28&  8.34& B9IV     &-0.05& 0.84& 270&0.009\\
8 & HD76649   & 08 55 50.4& -46 20 30&  8.33& B7II/III & 0.14& 0.67& 150&0.008\\
9 & HD76744   & 08 56 18.2& -46 19 57&  8.69& A0V      & 0.08& 0.75& 200&0.009\\
10& CD-454590 & 08 49 35.5& -46 23 18&  9.58& B5       & 0.20& 0.42& 140&0.010\\
11& CD-454606 & 08 50 15.0& -45 31 22&  8.96& B0.5V    & 0.38& 0.82& 180&0.009\\
12& CD-454645 & 08 51 34.9& -46 09 54& 10.32& A0       & 0.20& 0.14&  90&0.012\\
13& CD-454676 & 08 53 22.0& -46 02 09&  8.93& B0.5III  & 0.77& 0.35& 140&0.014\\
14& CD-464666 & 08 50 44.3& -46 38 11&  9.81& A0II     & 0.60& 0.40&  90&0.009\\
\hline                                                            
\end{tabular}
\end{minipage}
\end{table*}

RX J0852.0-4622 (Vela Jr., G266.2-1.2) is a young galactic SNR detected by means of its emission in hard
X-rays \citep{1998Natur.396..141A}, the $^{44}$Ti 1.16~MeV $\gamma$-ray
line \citep{1998Natur.396..142I}, radio \citep{2000A&A...364..732D}, and TeV
$\gamma$-rays \citep{2005A&A...437L...7A}. Vela Jr. with a 
diameter of $\sim2^\circ$ is superimposed on the eastern part of the well-known old
Vela SNR. The age and distance of Vela Jr. are estimated to $700\pm150$ yr
and $\sim200$ pc, respectively \citep{1999A&A...350..997A, 2005ApJ...632..294B}. 
The age, distance, and angular radius imply an average expansion velocity of 
$\sim5000$\kms.

Using {\em XMM-Newton}
images, \citet{2008ApJ...678L..35K} measured the proper motion of the bright NW rim of RX J0852.0-4622. 
The derived value turns out to be about 5 times lower than the predicted 
average expansion rate of Vela Jr. On the basis of the measured proper motion, 
\citet{2008ApJ...678L..35K} 
estimate the age of Vela Jr. to be 1700-4300~years and its
distance to be $\sim$750~pc. Although the conclusion could be hampered by a
possible interaction with the recently encountered dense interstellar medium,
the possibility of a large age cannot presently be fully ruled out.

Given the uncertainty in the age issue, we should carefully study the
implications of a young age. In this respect, it is tempting to consider the
unshocked ejecta of Vela Jr. using absorption spectroscopy against background
stars. Our preliminary estimates showed that broad Ca~II absorption lines might
be observed. Here we report results of spectral observations and analysis of the
spectra of distant stars across the Vela Jr. In Sect.~2, we describe the
observations and  data reduction. The results of the spectral synthesis and
extraction of broad and interstellar Ca~II absorption are presented in
Sect.~3. We fail identify any broad absorption lines and the implications of this are discussed in Sect.~4.

\begin{figure}[!b]
\includegraphics[width=70mm]{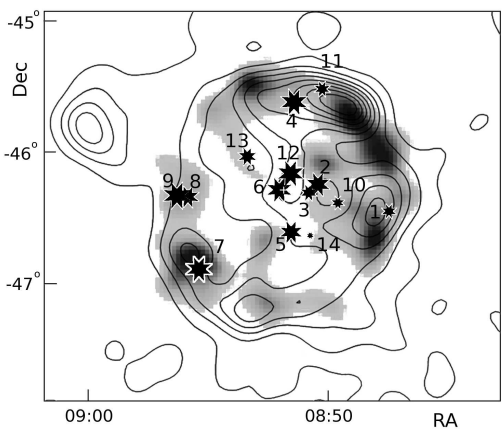}
\caption{Positions of the stars across the SNR images taken in TeV 
$\gamma$-rays (grayscale) and in hard X-rays (contours). The star 
numbers correspond to those in the first column of Table~\ref{tab:list}.
The cross is the position of the neutron star AX~J0851.9-4617. The size of
symbols qualitatively reflects the distance with the largest symbol being 
closest star.}
\label{fig:pos}
\end{figure}

\section{Observations and data reduction}

Spectra were obtained on the ESO 3.6-m telescope NTT 
(program 080.D-0012(A) PI: A.F.Iyudin) using the echelle spectrograph EMMI in 
BLMD mode of medium-dispersion spectroscopy with holographic grating~\#11
(3000~grooves/mm and maximum light reflectivity at $\lambda_{blaze}$=3500\AA).
The dispersion is 0.15\AA/pix, the resolving power is
$\lambda/\Delta\lambda$=9000 for a slit width of 1.02$''$ and a registered 
wavelength band of $\sim$1500\AA. The signal-to-noise ratio (S/N) is between 90
and 350. For each star, two overlapped spectra in the range of 3740-4021\AA\ 
were obtained with centers at 3820\AA\ and 3945\AA. This spectral 
region encompasses resonance spectral lines of Fe~I (3860\AA) and Ca~II
(3933\AA, 3968\AA). 

Among the observed stars, ten are of B-type and four of A-type (Table~\ref{tab:list})
with magnitudes $7^m<V<11^m$. Table~\ref{tab:list} contains 
the star number, star HD/CD name, right ascension, declination, $V$-magnitude,
spectral type, $B-V$ color index, relative impact parameter $p$, and
signal-to-noise ratio ($S/N$), and 1$\sigma$, which represents the root mean square error
in the spectral fit (see Sect.~3). The impact parameter is defined as
$p=\theta/\rho$, where $\theta$ is the angular distance of the star from the SNR
center and $\rho=1^\circ$  is the SNR angular radius. The SNR center presumably
coincides with the neutron star candidate AX~J0851.9-4617
\citep{2001ApJ...548..814S}, which has coordinates (RA,Dec)$_{2000}$=(08$^h$ 51$^m$
57$^s$, $-46^\circ 17'.4$). The preliminary estimated distances of selected
stars are in the range between 240-2000~pc, so at least some of the stars
lie behind Vela Jr. Positions of all stars of the program across the Vela Jr.
X-ray image (Rosat All Sky Survey data with energies above 1.3~keV) and TeV
$\gamma$-ray image \citep{2005A&A...437L...7A} are shown in Fig.~\ref{fig:pos}. 

Preliminary processing of the spectral CCD images, extracting the spectra, 
and performing wavelength calibration using ThAr-lamp data were all completed by employing the EMMI native {\it
emmi\_quickred} script of MIDAS; the atmosphere and interstellar extinction was also
taken into account for each individual star. The spectra are flux calibrated
with the help of spectra of the standard star HD~60753 taken in the same set of
observations. This star has a calibrated flux in the UV band only. The visual
flux is synthesized applying Kurucz ATLAS9 code~\citep{1993KurCD..13.....K} and
SynthVb package \citep{2003IAUS..210P.E49T}. A stellar atmosphere model is
calculated for solar metallicity, effective temperature \Teff=16200~K
($\sigma$=150~K), gravity \lgg=3.56 ($\sigma$=0.05), rotation velocity
\Vsini=24~\kms ($\sigma$=2~\kms), and interstellar absorption $A_V=0.25$
($\sigma$=0.03). The derived response function is used to produce flux-calibrated stellar spectra. One should emphasize that this procedure provides us
with the precise relative flux across a wide wavelength range but not the absolute
flux; the latter, in fact, is not needed for our purposes.

\begin{figure}[!b]
\includegraphics[width=80mm]{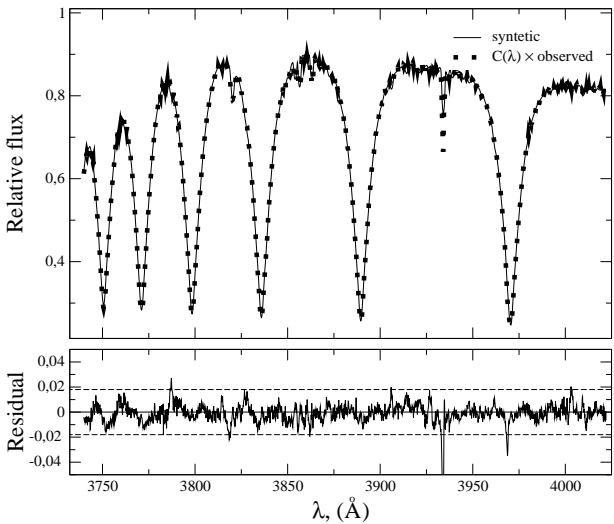}
\caption{Observed spectrum of the star HD~75968, synthetic spectrum, and 
the residual spectrum with 3$\sigma$ error box}
\label{fig:synth}
\end{figure}

\begin{table*}[!th]
\begin{minipage}{180mm}
\caption{Parameters of the stars and distances calculated 
by spectral method and from Hipparcos parallax}
\centering
\begin{tabular}{@{}lrcrrrccrr}
\hline\hline
\header{Star}
&\header{\Teff}&\lgg&\header{\Vsini}&\header{$V_r$}&\header{$(B-V)_0$}
&$A_V$&\header{M}&\header{$d_{sp}$}&\header{$d_{HIP}$}\\
&\header{(K)}&&(\kms)&\header{(\kms)}&\header{(mag)}&\header{(mag)}&\header
{ ($M_\odot$) } &\header{(pc)}&\header{(pc)}\\
\hline
HD75309  & 26500$\pm$400& 3.60$\pm$0.10&210&43& -0.25 &  0.8&  16 &
1900$\pm$300 & 
           \\                                   
HD75820  & 11400$\pm$200& 4.00$\pm$0.10&200&31& -0.10 &  0.2&  3.2& 
470$\pm$100 & 
505$\pm$184\\                                   
HD75873  &  8900$\pm$200& 2.50$\pm$0.05&15 &41&  0.01 &  1.2&  6  &
1400$\pm$200 & 
           \\                                   
HD75955  & 10400$\pm$200& 3.85$\pm$0.10&190&24& -0.07 &  0.2&  3.0& 
320$\pm$~70 & 
262$\pm$~35 \\                                  
HD75968  & 12250$\pm$150& 3.86$\pm$0.05&80 &35& -0.11 &  0.0&  3.8& 
570$\pm$140 & 
719$\pm$254\\                                   
HD76060  & 13400$\pm$200& 4.10$\pm$0.10&240&36& -0.13 &  0.1&  3.6& 
390$\pm$~90 & 
335$\pm$~63 \\                                  
HD76589  & 11800$\pm$200& 4.10$\pm$0.10&95 & 0& -0.10 &  0.1&  3.2& 
390$\pm$~90 & 
240$\pm$~76 \\                                  
HD76649  & 13300$\pm$150& 3.65$\pm$0.05&33 &33& -0.13 &  0.8&  4.5& 
640$\pm$110 &            \\
HD76744  & 10500$\pm$200& 4.20$\pm$0.10&150& -7& -0.08 &  0.5&  2.4& 
270$\pm$~50 &            \\
CD-454590& 22400$\pm$400& 3.60$\pm$0.10&140&31& -0.22 &  1.3&  11 &
2400$\pm$300 &            \\         
CD-454606& 29500$\pm$500& 3.80$\pm$0.10&240&33& -0.27 &  2.0&  17 &
1670$\pm$160 &            \\
CD-454645&  8400$\pm$200& 4.35$\pm$0.10&80 &35&  0.08 &  0.4&  1.8& 
330$\pm$~70 &            \\
CD-454676& 29000$\pm$500& 3.70$\pm$0.10&125&20& -0.27 &  3.2&  18 &
1080$\pm$150 &            \\
CD-464666& 10500$\pm$200& 2.05$\pm$0.05&30 &49& -0.08 &  2.1&  12 &
5700$\pm$500 &            \\
\hline
\end{tabular}
\label{tab:param} 
\end{minipage}
\end{table*}

\section{Analysis of spectra}

The obtained stellar spectra were analyzed by applying a synthetic flux
calculation employing the SynthVb and ATLAS9 codes 
to derive the stellar parameters \Teff, \lgg, and \Vsini\ (Table~\ref{tab:param}). 
To determine the rotation velocity, we use a standard method based on 
Fourier transformation of profiles of weak spectral
lines \citep{1933MNRAS..93..478C}. We adopted the limb darkening parameter 
$\epsilon=0.6$. 

The parameters of stellar atmosphere models were found by fitting the 
synthesized profiles to six Balmer lines (H7-H12). The fitting procedure
includes multiplication of the observed spectra by a fitting factor $C(\lambda)$,
which is defined as the linear approximation to the ratio of synthetic to observed 
spectrum $F_l^{syn}/F_l^{obs}$. In this case, we do not introduce any 
nonlinear distortion into the observed spectrum. The best fit is attained 
by minimizing the relative residual flux between the observed and synthetic spectrum 
\begin{equation}
r = \frac{C(\lambda)*F_l^{obs}-F_l^{syn}}{F_c^{syn}}\,, 
\end{equation}
where $F_c^{syn}$ is the continuum flux of the synthetic spectrum.
Using the relative flux permits us to avoid problems related to a 
poorly defined observed continuum. An example of the fit quality 
is demonstrated by Fig.~\ref{fig:synth}. The $1\sigma$ statistical
error in the spectral fit for individual stars are given in the last column of
Table~\ref{tab:list}. Any interstellar lines with an intensity exceeding the
$3\sigma$ error, which lie in the range of 1.5-4.2\%, should  be visible in the
residual spectra. As an example, the residual spectrum of star HD75968
(Fig.~\ref{fig:synth}) appeanrs to contain the Ca~II interstellar
absorption doublet. We believe that the systematic errors in producing
residual spectra are smaller than the quoted 3$\sigma$ statistical error.
\begin{figure}
\includegraphics[width=70mm,clip]{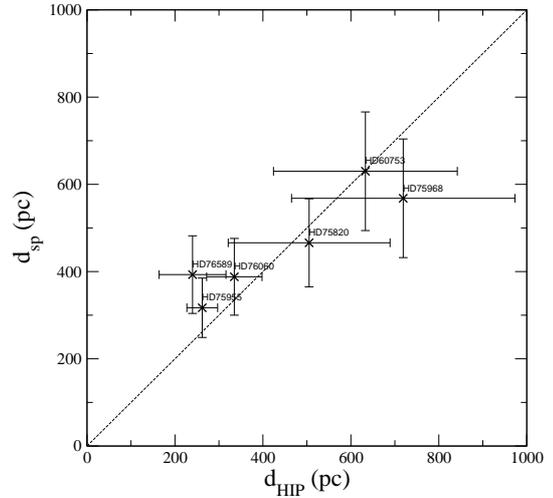}
\caption{Distance derived by the method of spectral parallax versus
distance according Hipparcos parallax.}
\label{fig:Rcmp}
\end{figure}

\subsection{Interstellar reddening and distance estimate}

Interstellar reddening in the observed spectra is taken into account using
extinction data reported by \citet{1990ARA&A..28...37M}. In the first 
step, we determine stellar parameters and calculate normal color indexes
$(B-V)_0$ \citep{2003IAUS..210P.A20C}. We then use the observed $(B-V)$ to
derive color excess $E(B-V)$ and $V$-band absorption, $A_V=3.1E(B-V)$. At the
second step, we improve our determination of the stellar atmosphere parameters and recalculate the
reddening.

To determine distances, we employ a modified method of spectral parallaxes in which
the stellar luminosity is derived from stellar evolutionary tracks as follows.
Using stellar parameters (\Teff\ and \lgg) and evolutionary
tracks \citep{1992A&AS...96..269S}, we estimate the stellar mass and thus derive
the bolometric luminosity. The absolute magnitude $M_V$ is then determined 
using the bolometric correction taken from \citep{1998A&A...333..231B}. The
luminosity and distance are thus determined from the standard formulae
\begin{equation}
\log\,L=-10.607+\log\,(M/M_\odot)+4\log\,T_{eff}-\log\,g\, ,
\end{equation}
\begin{equation}
M_V=4.69-2.5\log\,L-BC_V \,,
\end{equation}
\begin{equation}
\log\,d_{sp}=0.2(M_V-m_V-5+A_V)\,.
\end{equation}

\begin{figure*}[ht]
\centering
\includegraphics[width=170mm,clip]{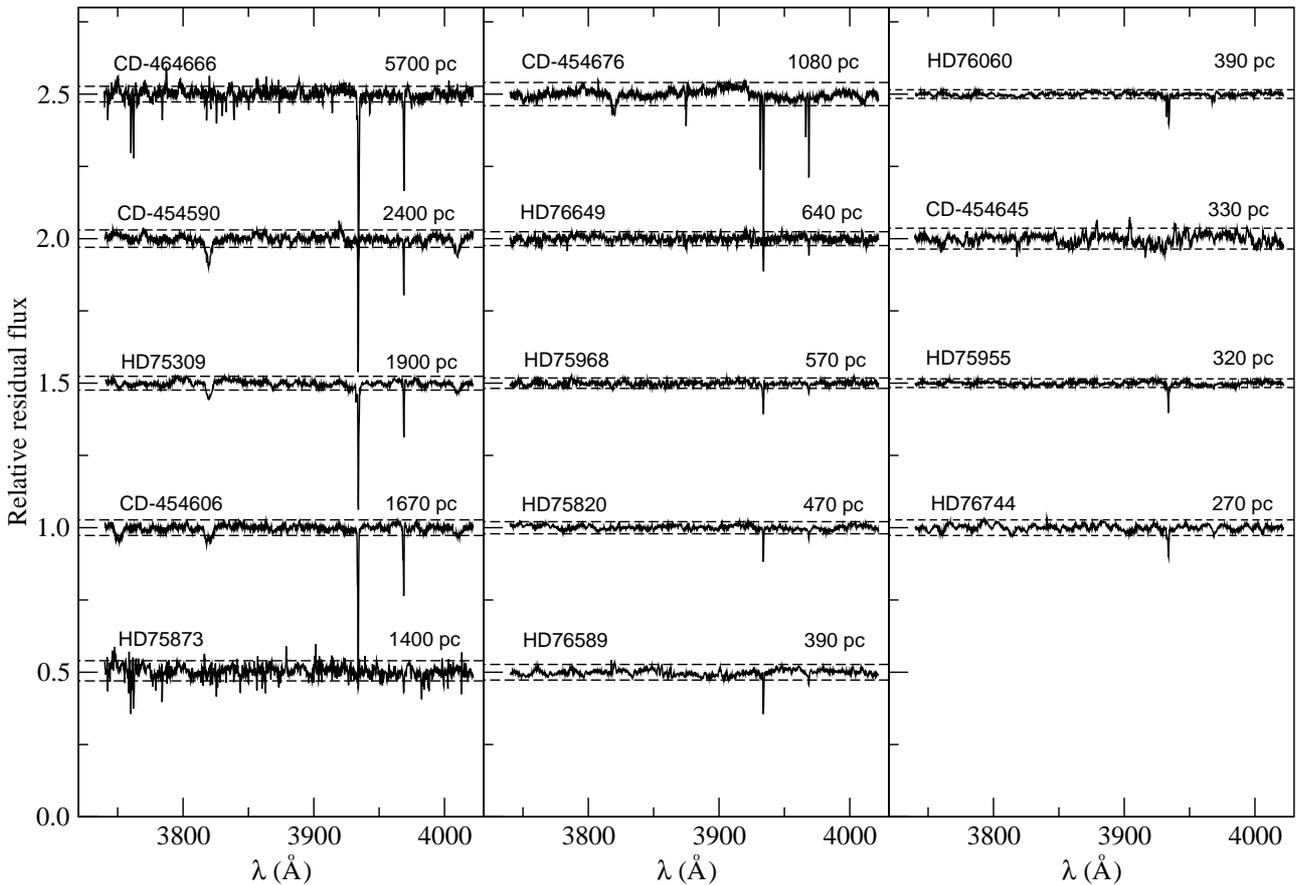}
\caption{The relative residual spectra with dashed lines showing the 3$\sigma$ statistical error boxes varying from $\pm0.015$ to $\sim$  $\pm$0.04 from star to star (for
1$\sigma$ errors in each case, see Table~\ref{tab:list}). The distance to each star is also indicated.}
\label{fig:res_all}
\end{figure*}
The normal color index $(B-V)_0$, interstellar absorption $A_V$, stellar 
masses $M$, and  distances are listed in Table~\ref{tab:param}. In the 
last column, we also indicate the available distances according to HIPPARCOS
parallaxes \citep{2007A&A...474..653V}. The distances determined by both methods
agree within errors (Fig.~\ref{fig:Rcmp}), which supports the reliability of
distances obtained by the method of spectral parallax.

\subsection{Interstellar lines}

The relative residual spectra for all the stars are shown in 
Fig.~\ref{fig:res_all}. The 3$\sigma$ levels are also indicated for each star. Large
values of 3$\sigma$ are seen for late B and A type stars with low rotation speeds
due to problems in fitting narrow stellar spectral lines using the accepted solar
abundance of chemical elements. The spectra do not uncover broad absorption 
resonance lines of Ca~II 3933\AA, 3968\AA, or Fe~I 3860\AA. 
In particular, the relative depth of the broad Ca~II absorption (if any) produced
by Vela Jr. is smaller than 0.04 at the level of $3\sigma$. We note, that the weak
absorption at 3819\AA\ and 4009\AA\ in the hottest stars of our sample are
related to helium lines, which are generally affected by non-LTE excitation and
cannot be modeled reliably in the LTE approximation. 

With the exception of two stars (HD~75873 and CD-454645), all the spectra contain
narrow unresolved interstellar Ca~II lines; their heliocentric radial velocities
are given in Table~\ref{tab:caii}. The interstellar absorption lines can be divided
into two major groups: low velocity $|V|$$\lesssim$50~\kms, and high velocity
$|V|$$\gtrsim$100~\kms. Most stars have one component with a positive radial
velocity of $\sim$22-48~\kms. The heliocentric velocity can be translated into a
LSR velocity in this direction using the relation $V_{LSR}=V_{hel}-13$~\kms.
This suggests that the dominant population of interstellar clouds in this
direction at the distances not exceeding 2 kpc, are characterized by the positive LSR velocities
$\sim$9-35~\kms. Two stars have negative low velocity components, of -13~\kms\ and -46~\kms. 

Three stars have high velocity 
components: HD~75309 (+153~\kms, -92~\kms), HD~76060 (-92~\kms), and 
CD-454676 (-150~\kms). 
These velocities are typical of high-velocity interstellar Ca~II
absorption found earlier in the direction of Vela SNR \citep{2000ApJS..126..399C}.
Interestingly, the spectrum of the star HD~75309 from our program was also
 studied by \citet{2000ApJS..126..399C}. Benefitting from the spectrum's high resolution, 
these authors were able to find eight components including 
 two high velocity components, +136~\kms\ and -107~\kms, and a strong low velocity
component +20~\kms. The corresponding interstellar absorption lines in
Table~\ref{tab:caii} are shifted redward by $\approx+16$~\kms, which reflects
the systematic difference in radial velocity between the two sets of data.
This may be partially related to the low resolution of our spectra. We studied
other sources of errors but were unable to explain this disparity. 

At least one star, CD-454676, exhibits conspicuous CN absorption  of electronic
transitions $R(0)$, $R(1)$, and $P(1)$  with the wavelengths of 3873.994\AA,
3874.602\AA, and 3875.759\AA, respectively. The heliocentric radial velocity of
these lines is +23~\kms\  ($V_{LSR}=+10$~\kms), which is consistent with the radial
velocity of Ca~II interstellar lines in the same star (Table~\ref{tab:caii}). 
 The equivalent width of $R(0)$ and $R(1)$ lines ($W(0)=0.01$\AA\ and
$W(1)=0.02$\AA) can be used to estimate the excitation  temperature  of the
rotational level $J=1$ and the column density of CN residing on  rotational
levels $J=0$ and $J=1$ in the weak line limit. Using  available oscillator
strengths of these transitions \citep{2002A&A...389..993G}, we obtain $T=6.8$~K
and column densities $N(0)=1.34\times10^{13}$~cm$^{-2}$ and 
 $N(1)=1.16\times10^{13}$~cm$^{-2}$. Assuming a Boltzmann population  of the
$J=2$  rotational level  $N(2)=5N(0)\exp(-16.325/T)$, we obtain  the total CN
column density $N=3.11\times10^{13}$~cm$^{-2}$. These values are comparable to
those of CN absorbers towards the Vela OB association \citep{2002A&A...389..993G}.

Two stars, HD~75873 and CD-454645, do not exhibit interstellar Ca~II lines. In
HD~75873 for which $d_{sp}$=1400~pc and $A_V$=1.2, the expected contribution of the
interstellar line to the equivalent width should be about 20\%. The explanation
of the apparent absence of absorption is the low rotation velocity
\Vsini=15~\kms\ and the high strength of stellar Ca~II absorption. Both factors
prevent us from distinguishing interstellar lines in this case. The second star,
CD-454645, at a distance of 330~pc is not expected to have strong
interstellar Ca~II lines. Given its very strong stellar Ca~II absorption, the
extraction of weak interstellar absorption in this case is precluded.

\begin{table}
\caption{Velocities of components of Ca~II interstellar absorption.}
\begin{tabular}{lcccccc}
\hline\hline
Star     &\multicolumn{2}{c}{$3933\AA$}&\multicolumn{2}{c}{$3968\AA$}\\
         &$\lambda (\AA)$& $V$ (\kms)&$\lambda (\AA)$& $V$ (\kms)\\
\hline
HD75309  &3935.67&  153&     3970.53&  156\\
         &3934.13&   36&     3968.91&   33\\
         &3933.06&  -46&     3967.84&  -48\\
         &3932.46&  -92&     3967.23&  -94\\

HD75820  &3933.98&   25&     3968.76&   22\\

HD75873  &3933.98&   25&     3968.76&   22\\

HD75955  &3933.95&   22&     3968.73&   19\\

HD75968  &3933.99&   25&     3968.76&   22\\

HD76060  &3934.14&   36&     3968.76&   22\\
         &3932.46&  -92&     3967.23&  -94\\

HD76589  &3934.10&   33&     3968.88&   31\\

HD76649  &3934.14&   36&     3968.76&   22\\

HD76744  &3933.95&   22&     3968.73&   19\\

CD-454590&3934.10&   33&     3968.88&   31\\

CD-454606&3933.11&   34&     3968.88&   31\\
         &3933.50&  -13&     3968.27&  -15\\

CD-454645&3934.29&   48&     3969.07&   45\\

CD-454676&3933.99&   25&     3968.76&   22\\
         &3931.70& -150&     3966.47& -151\\

CD-464666&3934.20&   41&     3968.87&   30\\
\hline
\end{tabular}
\label{tab:caii} 
\end{table} 

\begin{figure}
\includegraphics[width=85mm,clip]{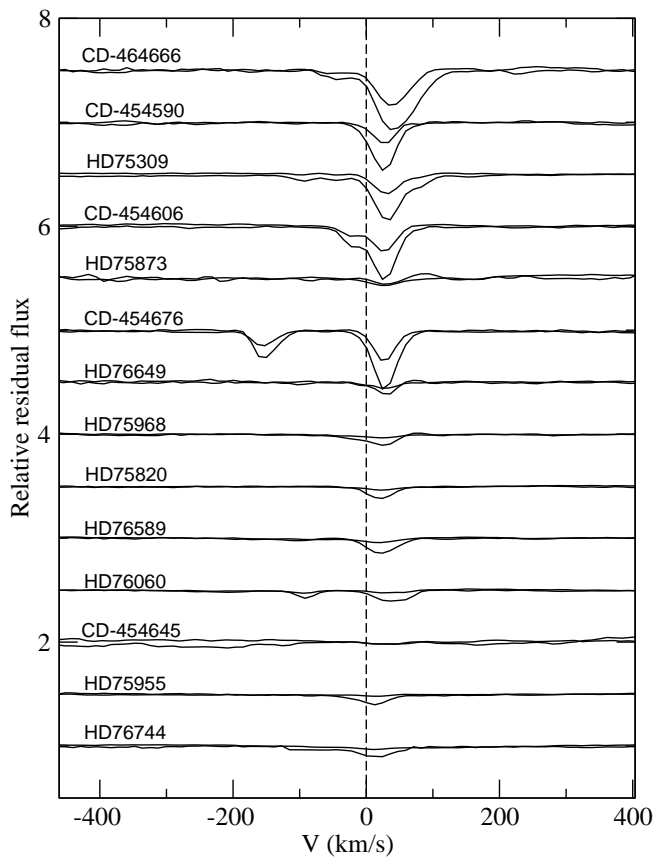}
\caption{The relative residual spectra for the stars. The distance increases 
upward.}
\label{fig:ca}
\end{figure}

%====================================================
\begin{table}
  \caption{Adopted parameters of supernovae}
  \begin{tabular}{ccccc}
\hline\hline
Parameters       &  SN~Ia & SN~IIP &  SN~Ibc & SN~Ic(h)  \\ 
\hline

 $M$ $(M_{\odot})$  & 1.4    &  10     &  3    & 4 \\
 $E$ $(10^{51} erg)$ & 1.4   &  1    &  1.5  & 20 \\ 
 
\hline
\end{tabular}
\label{t-snpar} 
\end{table} 
%============================================================

\section{Discussion}

\subsection{Narrow interstellar lines}

Most low velocity interstellar Ca II absorbers with $V_{LSR}\leq20$~\kms\ 
are most likely produced by local background clouds. This is also true for CN
absorbers, which in terms of velocity coincide with the Ca II absorbers. The large scatter in
velocities of between -50 and +30~\kms\, exceeding the usual dispersion in
cloud
velocities of $\sim 10$~\kms\, suggests that at least some of these absorbers are
related to clouds accelerated by either shock waves driven by wind bubbles of hot
stars or old SNR. This conjecture is in accord with results of observations
of interstellar ultraviolet O~I and Si~II absorption lines
\citep{1995ApJ...455..590W} towards the Vela SNR. Some of these lines  arise
from excited fine-structure levels in the Vela SNR direction  and indicate the
high pressure of clouds, $p\sim10^{-10}$~dyn~cm$^{-2}$
\citep{1995ApJ...455..590W}, which is two orders of magnitude higher than the average
pressure in the interstellar medium (ISM). 

The high velocity clouds with $|V_{LSR}|\sim100-150$~\kms\  are probably related
to the interstellar clouds shocked by the  expanding Vela SNR. Similar high
velocity gas was observed in  Ca II lines and ultraviolet lines corresponding to different
ions, including   C~I, O~I, Mg~I, and Mg~II \citep{1995ApJ...440..227J}, and
attributed to radiative shocks driven by the Vela SNR into interstellar clouds of
density $\sim10$ cm$^{-3}$. Interestingly, the closest star with a
high velocity interstellar component, HD~76060, lies at the distance
335$\pm$63~pc. This immediately provides an upper limit to the distance of the
Vela SNR of $\sim$335$\pm$63~pc, which is consistent with the Vela pulsar distance
of 294~pc \citep{2001ApJ...561..930C}.

\subsection{Broad absorption related to Vela Jr.}

The absence of Fe~I and Ca~II broad absorption lines in stellar spectra towards 
Vela Jr. requires
explanation. As we will see below, singly ionized metals should be more abundantly present in the Vela Jr. ejecta, 
and therefore absorption by neutral iron, with its relatively low ionization fraction and low 
value of the oscillator strength, should be significantly 
weaker than Ca II absorption.
We therefore concentrate on the absence of broad Ca~II lines. 
At least four possibilities are conceivable: Vela Jr. is farther 
than the most distant star in our sample; the SNR is much older; and the
Ca~II ionization fraction is small, i.e., Ca resides predominantly in the Ca~I 
or in the Ca~III ionization state. 
Discussion of these possibilities requires modeling the broad 
Ca~II absorptions expected at the given age for the remnants of the different SN types. 

\subsubsection{Broad Ca\,II absorption for different supernova types}

The unshocked ejecta expands freely, i.e., the expansion law at a 
given age $t$ is $v=r/t$. To describe this we will use cylindrical 
coordinates ($z$, $p$, $\phi$) with $z$-axis coincident with the line 
of sight directed towards the center of SNR.
The absorption produced by the scattering of the background stellar 
radiation in Ca\,II 3933, 3968\AA\ lines at the radial velocity 
$v_z$ is determined by a Sobolev optical depth 
in the resonant plane $z=v_zt$ along the line of sight of 
impact parameter $p$ (assuming that the ejecta is spherically symmetric)
\begin{equation}
\tau(z,p,\phi)=0.0265f_{12}\lambda_{12}n_1t\,,
\end{equation}
where the multipliers in the right-hand side in order are 
oscillator strength, wavelength, Ca\,II number density at the given 
radius $r=(z^2+p^2)^{1/2}$, and the SNR age $t$. 

For a given concentration of Ca determined by the density and Ca abundance,  
the line strength depends on 
the ionization fraction of Ca\,II. At the early phase of the ejecta expansion, at the time of $t\sim2$ yr after SN explosion, 
the calcium ionization in ejecta of any type SN 
is controlled primarily by the ionization loss of fast electrons 
(Compton electrons and positrons) produced by 
the radioactive decay chain $^{56}$Ni -- $^{56}$Co -- $^{56}$Fe and radiative 
recombination. At the stage of $t\geq3$ yr, spectra of SNe of 
different types are dominated by the emission lines of singly ionized 
metals, which indicates that singly ionized metals dominate. 
This is supported by numerical models of ionization and thermal 
balance of ejecta powered by the radioactive decay of $^{56}$Co 
for  SN~Ia \citep{1980PhDT.........1A} and SN~IIP \citep{1998ApJ...496..946K}.
At the later stages $t\sim10$ yr, the ionization is dominated by positrons from
$^{44}$Ti decay. Our estimate indicates that even a maximal expected mass
$10^{-4}~M_{\odot}$ of $^{44}$Ti is insufficient to maintain the high ionization
of Ca. At later stages, therefore, recombination 
dominates and Ca may become mostly neutral. 

However, at the SNR age of $t\sim10^2$ yr the characteristic recombination time of Ca 
is larger than the expansion time. At this stage, the ionization 
by the  starlight may become essential.
For example, \citet{1999ApJ...514..195F} demonstrate that calcium in the ejecta 
of SN~1885 in M31 
can be efficiently ionized by the bulge starlight within the ionization 
time of 
about 10 yr. In the case of Vela Jr. we use the model of the starlight spectrum 
in the Galactic plane at the radius of 7.5 kpc given 
by \citet{2006ApJ...648L..29P}, 
and the photoionization cross-sections of \citet{1996ApJ...465..487V}. The found
photoionization time is $\sim180$ yr for Ca\,I and $\sim3\times10^3$ yr for
Ca\,II.  At the age of $\sim700$ yr, we thus expect that Ca in Vela Jr. 
should be singly ionized.

%======================================================
\begin{figure}
\includegraphics[width=90mm]{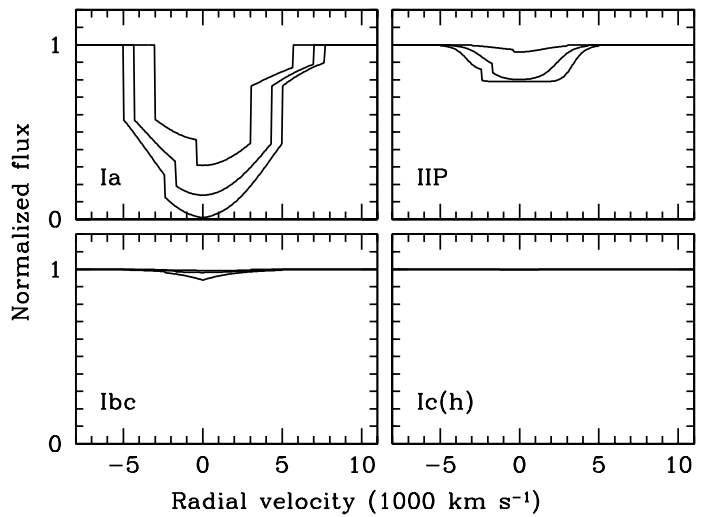}
\caption{
Absorption profile of Ca\,II doublet expected in the stellar spectrum 
of different progenitors of Vela Jr. Shown are cases of impact parameter 
equal to 0, 0.5, and 0.8. The strongest absorption always corresponds 
to zero impact parameter.
}
\label{f-prof}
\end{figure}
%==========================================================

The ionization of Ca\,II by X-rays from the reverse shock, and by accelerated 
protons may also play a role. The X-rays ionize metals from $K$ and 
$L$ shells, 
and photoelectrons then ionize Ca\,II. For the observed X-ray 
flux $f_{\rm x}\sim10^{-10}$ erg cm$^{-2}$ s$^{-1}$ in the 0.5-10 keV band, 
the characteristic photoionization time for Ca\,II at the SNR age of 700 yr 
is found to be $\sim10^8$ yr 
for SN~Ia, and this process is thus negligible.
Ionization by relativistic protons accelerated in the shock wave can be 
estimated by assuming an average efficiency of cosmic ray acceleration 
per SN of 10\%  and the shock wave energy 
of $\sim10^{51}$~erg. We find then that the ionization time for Ca\,II 
at the age of 700~yr is $\sim2\times10^3$~yr, which is larger than the SNR age.
We thus conclude that cosmic rays for the adopted acceleration efficiency 
essentially cannot ionize Ca\,II, so all the calcium in the unshocked 
ejecta of Vela Jr. is expected to remain in Ca\,II. 

The predicted profiles of the Ca\,II doublet at the age of 700~yr 
for different varieties of SNe are shown in Fig.~\ref{f-prof} assuming  
that all the calcium is in the Ca\,II state.  We assume that in SN~IIP and SN~Ibc 
the Ca abundance is solar, while for SN~Ia we assume that the Ca/Fe ratio 
by mass is solar, while the total mass of iron in the ejecta is $0.6~M_{\odot}$.
Ejecta parameters for different SNe are given in Table~\ref{t-snpar}. 
Apart from SN~Ia, SN~IIP, and SN~Ibc, we also consider energetic SN~Ic, so-called 
hypernovae, which is designated hereafter as SN~Ic(h).
The boundary velocity of the unshocked ejecta is taken to be 5000~\kms
~in accordance with the distance of 200~pc and the age of 700~yr.
The density distributions $\rho(v)$ in the unshocked ejecta are assumed to be 
exponential for compact pre-SNe and to form a plateau with the outer power law 
$\rho\propto v^{-9}$ for SN~IIP. The plotted profiles are computed for 
three values of impact parameter in units of the angular radius: 0, 0.5, and
0.8. The absorption is predicted to be deep for all the impact parameters 
in the case of  SN~Ia, rather deep for SN~IIP, very weak for SN~Ibc, and 
negligible (relative depth $<0.006$) in the case of SN~Ic(h). 
If the age and distance of Vela Jr. are close to the values adopted above, 
the progenitor would be unlikely of type SN~Ia or SN~IIP; instead an association of the SNR with 
a SN~Ibc or SN~Ic(h) is quite plausible. 

%======================================================
\begin{figure}
\includegraphics[width=90mm]{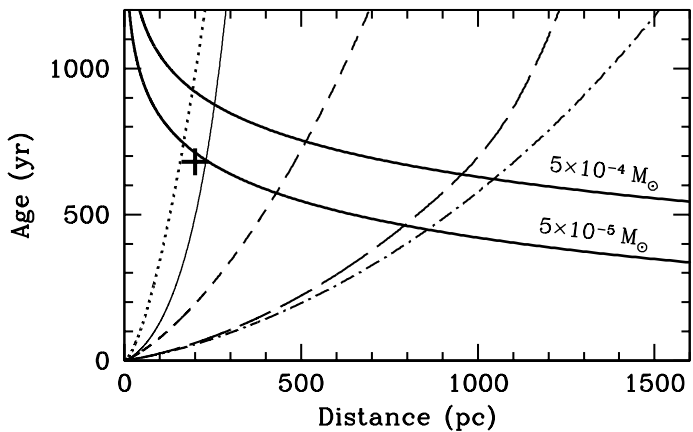}
\caption{
Age-distance relations provided by $^{44}$Ti mass (thick solid lines) 
and radius of the supernova remnant. The radius is calculated for SN~IIP 
(dotted line), SN~Ia (thin solid line), SN~Ibc (short-dashed line), 
SN~Ic(h) with $35~M_{\odot}$ progenitor (long-dash line), and $60~M_{\odot}$ 
progenitor (dashed-dotted line). Cross shows parameters derived
by \citet{1998Natur.396..142I}.
}
\label{f-aged}
\end{figure}
%==========================================================

\subsubsection{How far away might Vela Jr. be?}

We now relax arguments used earlier to constrain the age and distance 
of Vela Jr. \citep{1998Natur.396..141A, 1998Natur.396..142I} and 
check whether Vela Jr. lies at a very 
large distance, $>1$ kpc, beyond any star in our sample. 

For a given $^{44}$Ti mass, a combination of age and distance is constrained by
the observed flux in the gamma-ray line 1.16~MeV. Nucleosynthesis models predict a
production of $(1-5)\times10^{-5}~M_{\odot}$ of $^{44}$Ti in
SN~Ia \citep{1998Natur.395..672I} and $10^{-5}-10^{-4}~M_{\odot}$ in
core-collapse SNe \citep{1995ApJS..101..181W}. An independent estimate of the
$^{44}$Ti yield per SN can be obtained by assuming that almost all the
$^{44}$Ca is produced as $^{44}$Ti both in SN~Ia \citep{1999ApJS..125..439I} and
core-collapse SNe \citep{1995ApJS..101..181W}. The solar mass ratio of
$^{44}$Ca to $^{56}$Fe is $10^{-3}$, which means that SN~IIP and SN~Ibc
producing $0.05-0.1$ of $^{56}$Ni per SN should eject $5\times10^{-5}-
10^{-4}~M_{\odot}$ of $^{44}$Ti, while SN~Ic(h) producing, in a similar way to SN~1998bw,
up to $0.5~M_{\odot}$ of $^{56}$Ni \citep{1998Natur.395..672I} is able to eject
as much as $\sim5\times10^{-4}~M_{\odot}$ of $^{44}$Ti. We therefore, expect,
that the mass of $^{44}$Ti ejected by SNe of different types 
lies in the range $5\times10^{-5}-5\times10^{-4}~M_{\odot}$. The corresponding
relationship between the age and distance suggested by the observed flux of the
1.16~MeV line $3.8\times10^{-5}$ cm$^{-2}$~s$^{-1}$ is shown in
Fig.~\ref{f-aged} for the two extreme values of ejected $^{44}$Ti mass. 

Deceleration of supernova ejecta in the interstellar medium provides us
with another relation between the distance and age for a given choice of ejecta
parameters, ISM density, and angular radius of Vela Jr. We compute the
interaction of ejecta with the ISM in the thin shell approximation 
\citep{1982ApJ...259..302C} assuming typical ejecta
mass and energy (Table~\ref{t-snpar}). In the case of SN~IIP, the adopted
hydrogen number density of ISM is 0.3 cm$^{-3}$, which is the average density of the
warm neutral medium (WNM). The latter comprises about 80\% of the ISM
mass \citep{1995ApJ...443..152W}. For SN~Ia apart from WNM, we also consider the
ISM in the form of a hot ionized medium (HIM) of density
0.003~cm$^{-3}$. This gas occupies about 50-60\% of the
volume \citep{1995ApJ...443..152W}. As in the case of SN~Ibc and SN~Ic(h), they explode in
the ISM modified by the fast main-sequence wind, slow red supergiant wind, and
the Wolf-Rayet wind. We consider $35~M_{\odot}$ and $60~M_{\odot}$ as template
progenitor stars; both cases were explored by \citet{1996A&A...316..133G}.
According to these results the pre-SN in the $35~M_{\odot}$ case is imbedded in a
hot bubble of the uniform density of $\sim0.003$~cm$^{-3}$ with a radius of
$18$~pc surrounded by a dense cool shell of total mass
$\sim20~M_{\odot}$. For the $60~M_{\odot}$ progenitor, the bubble density is
$\sim0.001$~cm$^{-3}$ and its radius is $50$~pc. We note in passing that a 
model of Vela Jr. (RXJ0852.0-4622) taken to be of the SNII/Ib type exploded in a 
wind blown cavity was considered by \citet{2009A&A...505..641B} .

The age-distance relations for all the discussed cases are shown
in Fig.~\ref{f-aged}. The SN~Ia exploded in the HIM phase shows almost the same 
age-distance relation as SN~Ibc and is therefore not shown in this figure. This
diagnostic plot is similar to that used by \citet{1999ApJ...514L.103C}. The
essential difference, however, is that they used a set of arbitrary expansion
velocities of the swept-up shell, while we calculate the evolution of the shell
radius for each type of SN. For a given age, the minimal distance corresponds to
a SN~IIP expanding in the WNM phase, while the maximal distance corresponds to a
SN~Ic(h) with a $60~M_{\odot}$ progenitor. In combination with the $^{44}$Ti
curves, these two cases imply the allowed ranges of 450-900~yr and 150-1000~pc
for the age and distance of Vela Jr., respectively. The major result of this plot
is that the distance of Vela Jr. cannot exceed 1 kpc. We thus conclude that at
least several stars in our sample (Table~\ref{tab:param}) are behind the SNR.
This permits us to disregard the explanation of the absence of broad Ca~II
absorption being the result of the large distance to Vela Jr. 

\subsubsection{Was the progenitor of Vela Jr. a hypernova?}

The absence of broad Ca\,II absorption in the spectra of stars at distances
$>1$ kpc suggests that the SNR progenitor was either of SN~Ibc or SN~Ic(h)
because only for these SNe is the expected absorption weak and possibly undetected (Fig.~\ref{f-prof}). To distinguish between these two SN
possibilities, one should take into account the intrinsic width of the 1.16~MeV line. 

\begin{figure}
\includegraphics[width=90mm]{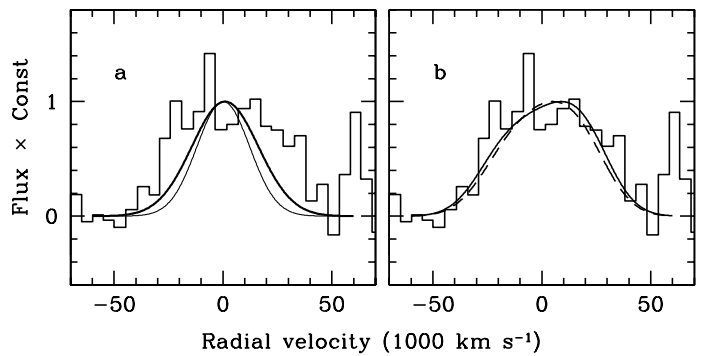}
\caption{ Profile of 1.16 MeV line of $^{44}$Ti. Panel {\em a}: profiles for
SN~Ibc (thin line) and SN~Ic(h) (thick line) with spherically-symmetric
distribution of $^{44}$Ti. Panel {\em b}: profiles for SN~Ic(h) with $^{44}$Ti
distributed in the external parts of bi-polar jets at inclination angles of
$30^{\circ}$ (solid line) and $60^{\circ}$ (dashed line).}
\label{f-tiprof}
\end{figure}

In the case of SN~Ibc at the age
of 650~yr, the expected profile of the $^{44}$Ti line (cf. Fig.~\ref{f-aged}) convolved with the instrumental profile
($\sigma=45$ keV) is found to be too narrow compared with the observed one
(Fig.~\ref{f-tiprof}a), even assuming homogeneous mixing of $^{44}$Ti up to a
velocity of 10 000~\kms. The observed broad profile of the 1.16~MeV line implies
that most of the ejecta mass consisting of $^{44}$Ti has significantly larger velocities. However, the SN~Ic(h)
case with maximal expansion velocities of 31 300~\kms at the age of $\sim500$ yr
(the case of $60~M_{\odot}$)  and spherically-symmetric distribution of
$^{44}$Ti homogeneously mixed to 31 000~\kms ~does not help resolve the ambiguity
either (Fig.~\ref{f-tiprof}a).

The solution to the line width problem might be found by taking into account
that iron-peak elements are ejected by SN~Ic(h) in the form of high velocity
bipolar jets and assuming that $^{44}$Ti resides only in the outer parts of the
jets. $^{56}$Ni-rich bipolar jets are predicted by the collapsar model
\citep{1999ApJ...524..262M} proposed for the hypernova SN~1998bw, and the jet-like
structure is consistent with the spectral line profiles
\citep{2006ApJ...645.1331M}. \citet{2003ApJ...598.1163M} also predict
an external location of $^{44}$Ti. In the 1.16~MeV profile simulations, we assume
that $^{44}$Ti is homogeneously distributed along the radius in the velocity
range of $20 000-31 000$~\kms ~within jets of opening angle $60^{\circ}$
and inclination angle $\theta$. We took into account the light travel-time delay
that produces the profile skewed towards red. Two cases are shown
(Fig.~\ref{f-tiprof}b) for the angle between the jet axis and the line of
sight,  $\theta=30^{\circ}$ and $\theta=60^{\circ}$, both of which fit the data more closely
than the spherically symmetric model. We therefore conclude that the hypernova
model with the outer location of $^{44}$Ti in bipolar jets of SN~Ic(h)
is consistent both with the absence of the broad Ca~II absorption, and the broad 1.16~MeV profile. 

A problem with the SN~Ic(h) scenario is that the high ejecta velocity
implied by this model infers a low ambient density, which seems to disagree
with the baryonic origin of TeV gamma-ray emission from Vela Jr.  Gamma ray production via
$pp$-collisions seems to be the preferred model compared to the inverse Compton
mechanism \citep{2005A&A...437L...7A, 2009A&A...505..641B}. An alternative may be provided by assuming
that we see the early stage of the interaction of the SNR with a
dense environment that has not yet been affected by the previous expansion dynamics.
This conjecture is in line with the low expansion velocity found for the NW rim by 
\citet{2008ApJ...678L..35K}. 
Another discomfort is related to the hypernova being a rare variety
of SNe that comprises only about 1\% of all SNe~Ibc
\citep{2004ApJ...607L..17P}. Only high signal-to-noise spectral imaging of
Vela Jr. in the 1.16~MeV line band with energy resolution of $\leq40$ keV and
angular resolution of $\leq1^{\circ}$ may confirm (or reject) the high velocities
of $^{44}$Ti and detect any jet-like (if any) structure of the $^{44}$Ti
distribution.

Given the difficulties arising in the interpretation of data on Vela Jr.,
we should not exclude out completely the possibility that this SNR is older
\citep{2008ApJ...678L..35K}. However, only additional observations at different
wavelength bands will be able to help pinpoint the age and the origin of Vela Jr.

%======================================================
%==========================================================

\section{Conclusions}

We have presented our attempt to detect unshocked ejecta of the young 
SNR Vela Jr. by analyzing broad Ca II absorption lines in spectra of background stars. We
obtained and analyzed spectra of 14 stars across Vela Jr. using standard
methods of spectral synthesis. We concluded that broad absorption lines are
absent. The $3\sigma$ upper limit to the depth of broad absorption lines is
0.04. We detected both low velocity and high velocity interstellar Ca~II 
absorptions. The latter are attributed to the cold gas of  radiative shocks
propagating in clouds engulfed by the old Vela SNR.

The absence of broad Ca II absorption lines and the constraints imposed by
the flux of the $^{44}$Ti gamma-ray line and the angular size of the SNR 
imply that only SN~Ibc or energetic SN~Ic (hypernovae) could have produced
Vela Jr., if our estimates of the age and distance are correct. The
additional constraint provided by the width of the 1.16 MeV  $^{44}$Ti
line also supports the hypernova scenario for Vela Jr. origin. However, we
emphasize the need for more reliable data on the $^{44}$Ti gamma-ray line
profile and higher resolution imaging of Vela Jr. in the gamma line to verify
the hypernova scenario.

\bibliographystyle{aa}
\bibliography{vela}

\end{document}